\documentclass[11pt]{article}
\usepackage{float}
\usepackage{amsmath}
\usepackage{amsthm}
\usepackage{amssymb}
\usepackage{amsfonts}
\usepackage{authblk}
\usepackage{graphicx}
\usepackage{booktabs}
\usepackage{multirow}
\usepackage{setspace}
\usepackage[T1]{fontenc}
\usepackage{indentfirst}
\usepackage[footnotesize,bf]{caption}
\usepackage[ruled,vlined]{algorithm2e}
\usepackage[round, authoryear]{natbib}
\usepackage{xcolor}
\usepackage[colorlinks,linkcolor=blue,citecolor=blue]{hyperref}
\usepackage{bm}
\def\R{\mathbb{R}}

\def\I{\mathbb{I}}
\def\E{\mathbb{E}}
\def\P{\mathbb{P}}

\def\cB{\mathcal{B}}

\def\cE{\mathcal{E}}

\def\cI{\mathcal{I}}

\def\cN{\mathcal{N}}

\def\cS{\mathcal{S}}

\def\spa{\mathrm{span}}
\def\Bin{\mathrm{Bin}}

\def\con{+}
\def\Var{\mathrm{Var}}
\def\Cov{\mathrm{Cov}}

\DeclareMathOperator*{\argmax}{arg\,max}

\theoremstyle{plain}
\newtheorem{theorem}{Theorem}

\newtheorem{proposition}{Proposition}

\theoremstyle{definition}
\newtheorem{definition}{Definition}

\newtheorem{assumption}{Assumption}

\theoremstyle{remark}
\newtheorem{remark}{Remark}

\counterwithin{equation}{section}
\usepackage{xcolor}

\title{A hybrid model for zero-inflated proportion data}

\date{}

\author[1]{Yangzi Zheng}
\author[2]{Qian Li}
\author[3]{Binyan Jiang}
\author[3]{Pinjun Dong}

\affil[1]{\small Beijing International Center for Mathematical Research,
	\small	 Peking University, China}

\affil[2]{\small School of Mathematics and Statistics, 
	\small Shaanxi Normal University, Shaanxi, China
}
\affil[3]{\small Department of Data Science and Artificial Intelligence, 
	
	\small The Hong Kong Polytechnic University, Hong Kong, China}

\begin{document}
\maketitle

\begin{abstract}
\noindent
This paper proposes a hybrid model for analyzing zero-inflated proportion data by distinguishing two types of zeros. 
The first type arises from sampling variability, whereas the second reflects intrinsically unsuitable covariates. 
Accordingly, the hybrid model combines a regression component for suitable observations with an estimator of the suitability region that identifies unsuitable covariates. 
Using a debiasing technique, we establish the consistency of both the regression and classification components, together with the asymptotic normality of the regression coefficient and a convergence rate for the support estimator.
Comprehensive simulation studies verify our theoretical findings.
We further apply the model to two real datasets, where it outperforms most competing methods.
\end{abstract}


\hypersetup{linkcolor=black}
\tableofcontents
\hypersetup{linkcolor=blue}
\newpage
\section{Introduction}\label{sec:introduction}

Proportion data with an atom at zero are common across the sciences. In ecology, aquaculture, and fisheries, one records fractions such as the proportion of germinating seeds, hatching eggs, or target-species catch, and these fractions are often zero. Such zeros arise from one of two mechanisms. One is a \emph{sampling zero}, an unfavorable realization of a positive success probability. The other is a \emph{structural zero}, in which the covariates are intrinsically unsuitable and no positive outcome could have been realized.

\begin{figure}[h]
\centering
\includegraphics[width=\textwidth]{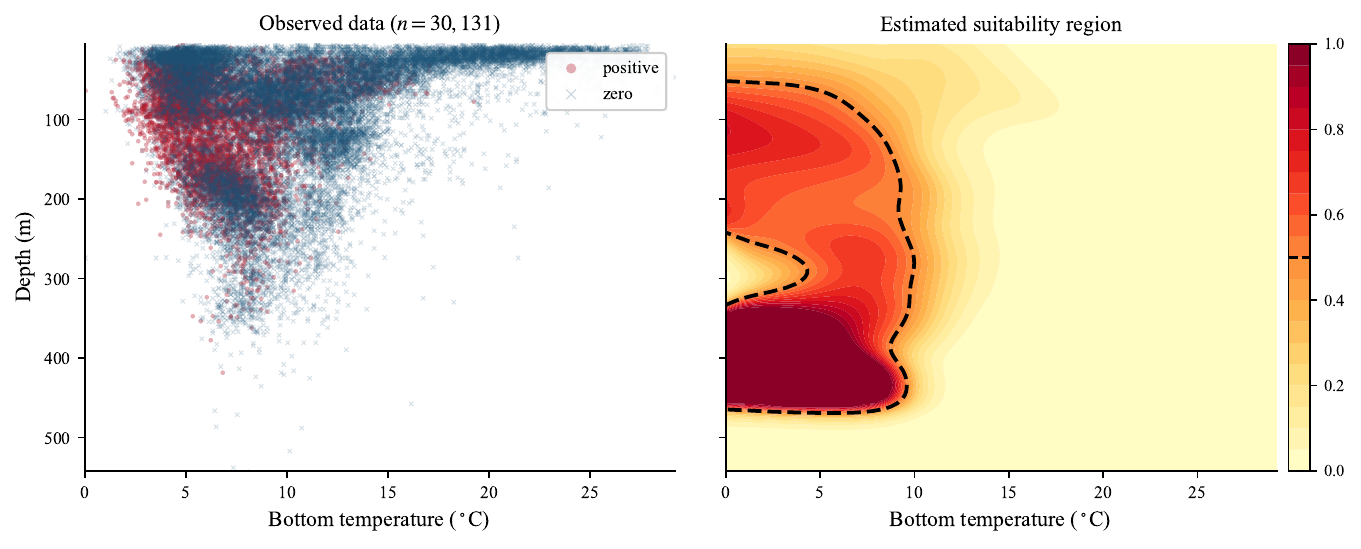}
\caption{Northeast US bottom-trawl survey. Left: observed data. Right: the suitable region recovered by our method.}
\label{fig:trawl-intro}
\end{figure}

One example comes from the Northeast US bottom-trawl survey \citep{flanagan2018response}. The survey recorded the proportion of Atlantic cod in the catch for $34{,}608$ hauls, with bottom temperature and depth available for $30{,}131$. In $2{,}958$ hauls beyond the $18^{\circ}$C bottom isotherm, only one cod was caught among a total of $8.8\times10^{7}$ fish, whereas colder waters yielded millions of cod. The positive-catch rate was higher at depths of $100$--$300$~m. The visualization reveals nonlinear patterns in the trawl results with respect to temperature and depth, and these patterns motivate us to propose a semi-parametric model~\eqref{eq:sampling} to analyze this phenomenon. Figure~\ref{fig:trawl-intro} visualizes the sampling results and the suitable region recovered by our model. The latter reflects the behavior of the zero structure in the survey.

\subsection{Related work}\label{subsec:related-work}

Existing approaches for zero-inflated proportion data can be broadly grouped into three frameworks. Within the frequentist framework, a common strategy is to model the data in two parts. For instance, two-part and hurdle models \citep{cragg1971some,mullahy1986specification,zeileis2008regression} first use a logistic regression to separate zeros from positive values, and then fit a truncated model to the positive magnitudes alone. Alternatively, zero-inflated models \citep{lambert1992zero,hall2000zero} use a latent mixing component, where one part generates excess zeros and the other generates counts or continuous values. For continuous proportions, zero-or-one inflated beta regression \citep{ferrari2004beta,ospina2012general} extends this mixture to handle boundary atoms, while fractional-response regression \citep{papke1996econometric} models only the conditional mean via quasi-maximum likelihood. Beyond the frequentist framework, \citet{ali2020constrained} adopts an optimization framework by imposing a constrained formulation on the marginal mean, and \citet{koslovsky2023bayesian} takes a Bayesian approach using latent Bernoulli at-risk indicators with smoothly modeled probabilities. However, these models all treat the structural zero as a probabilistic event, and thus do not target a deterministic structure for the zero outcome. In ecology, the occupancy--detection framework of \citet{mackenzie2002estimating} addresses a closely related zero dichotomy (nondetection vs.\ absence) but models occupancy smoothly and requires repeated visits. Our support estimator also draws on the kernel level-set literature \citep{cadre2006kernel,willett2007minimax}, which provides recovery guarantees for thresholded regression functions.

The zero-inflated binomial model of \citet{hall2000zero} assumes
\[
mY \mid X \sim \begin{cases} 
	0, & \text{with probability } \sigma(X^\top\alpha), \\
	\Bin(m, \sigma(X^\top\beta)), & \text{with probability } 1 - \sigma(X^\top\alpha).
\end{cases}
\]
Here, the mixing probability is strictly between zero and one, so the structural zero is treated as a smooth random event and the boundary of the unsuitable region is not sharply defined. Instead, we consider the structural zero component as a deterministic property of the covariates. Specifically, we model the outcome through
\[
mY\mid X\sim \Bin(m,\sigma(X^\top\theta^*)\I_{S^*}(X)).
\]
We call it a hybrid model, as it replaces the random mixing probability with an indicator function, making the suitability region a set \(S^*\) whose boundary can be explicitly targeted for estimation. More details can be found in Section~\ref{sec:methodology}.

\subsection{Contributions}\label{subsec:contributions}

This hybrid model raises questions of identification, estimation, and practical use. Our contributions are as follows.

\begin{itemize}
	\item We propose a hybrid model to capture the zero-inflated proportion data. To exclude the zero outcomes, we design a suitable region that characterizes where positive samples can occur. The identifiability of separating the traditional regression coefficients from the suitable region is rigorously established under an equivalent assumption.
	\item We estimate the coefficients using a debiased likelihood based on the positive observations and recover the region by reweighted kernel regression. The asymptotic properties of both the coefficient and the region provide theoretical guarantees for the estimation.
    \item We apply the proposed method to two real datasets. It consistently outperforms nearly all parametric models and is even on par with a tabular foundation model that is increasingly popular nowadays.
\end{itemize}

\subsection{Organization}\label{subsec:organization}

The rest of the paper is organized as follows.
Section~\ref{sec:methodology} introduces the model, its identifiability, and the debiased approach. Section~\ref{sec:estimation} develops the
estimation procedure. Section~\ref{sec:theoretical} presents the
theoretical results. Section~\ref{sec:numerical} contains simulations and
real data analyses. Proofs are in the Appendix.

\subsection{Notations}\label{subsec:notations}

Let $\R^d$ denote the $d$-dimensional Euclidean space, equipped with the inner product $\langle\cdot,\cdot\rangle$ and the induced norm $\|\cdot\|_2$.
We write $\P$ and $\E$ for probability and expectation, respectively, and $\I(\cdot)$ for the indicator function.
For sets $A$ and $B$, let $A\triangle B:= \{x : x\in A \text{ or } x\in B,\ x\notin A\cap B\}$ denote the symmetric difference. The boundary of $A$ is denoted by $\partial A$.

At the population level, we use $X\in\R^d$ to denote the feature vector and $Y\in[0,1]\subseteq\R$ to denote the proportional outcome, where $m$ is the number of trials. The realizations of $(X,Y;m)$ are denoted by $\{(X_i,Y_i;m_i)\}_{i=1}^n$, which form $n$ independent observations. The true regression coefficient and suitability domain are denoted by $\theta^*\in\R^d$ and $S^*\subseteq\R^d$, where the superscript $(\cdot)^*$ marks true values. Correspondingly, we use a hat $\hat{(\cdot)}$ to denote estimators, such as $\hat\theta$ and $\hat S$.
The parameter spaces are denoted by $\Theta$ and $\cS$, whose definitions are given in the main body of the paper.
Asymptotic behavior is often captured by $\to_{\mathrm P}$ and $\to_{\mathrm d}$, which represent convergence in probability and convergence in distribution, respectively.
The density function of $X$ is denoted by $f$, and $K_h$ denotes a kernel function with bandwidth $h$.
\newpage
\section{Methodology}\label{sec:methodology}

Let $(X,Y;m)$ denote an observation, where $X\in\R^d$ is a feature vector, $Y\in[0,1]\subseteq\R$ is a proportion-valued outcome, and $m$ is the number of trials, which may vary across observations. Suppose that
\begin{equation}\label{eq:sampling}
	mY\mid X\sim
	 \Bin\bigl(m,\sigma(X^\top\theta^*)\I_{S^*}(X)\bigr).
\end{equation}
Here, $\sigma(X^\top\theta^*)$ is a parametric component used to capture sampling zeros, where $\sigma(x)=[1+\exp(-x)]^{-1}$ is the sigmoid function. The suitability of the covariates is characterized by the indicator function $\I_{S^*}(X)$, so that a covariate vector $X$ is deemed unsuitable whenever $X\notin S^*$.

Model~\eqref{eq:sampling} involves two unknown components that need to be estimated: the regression coefficient $\theta\in\Theta:=\R^d$ and the suitability domain $S\subseteq\R^d$. However, the model is not identifiable (Definition~\ref{def:iden}) when $\theta\in\Theta$ and $S\subseteq\R^d$ are unrestricted. To address this, we restrict $S$ to the class of full-span domains in Definition~\ref{def:ssd}, which yields Proposition~\ref{prop:iden}.

\begin{definition}\label{def:ssd}
	We define $\cS$ as the class of \emph{full-span domains}, given by
	\begin{equation*}
		\cS:=\{S\subseteq\R^d: \dim(\spa(S))=d\}.
	\end{equation*}
\end{definition}

\begin{definition}\label{def:iden}
	We say that model~\eqref{eq:sampling} is identifiable
	on $\Theta\times\cS$ if, for any
	$(\theta,S),(\theta',S')\in\Theta\times\cS$,
	$\P_{\theta,S}(Y\mid X)\equiv
	\P_{\theta',S'}(Y\mid X)$ implies
	$(\theta,S)=(\theta',S')$.
\end{definition}

\begin{proposition}\label{prop:iden}
	The model~\eqref{eq:sampling} is identifiable on $\Theta\times\cS$ if and only if $S\in\cS$.
\end{proposition}

\begin{remark}
	In Definition~\ref{def:iden}, the conditional distribution
	$\P_{\theta,S}(Y\mid X)$ is viewed as a function of $X$.
	Identifiability therefore requires the map from
	$(\theta,S)$ to the induced conditional distribution
	function to be injective.
\end{remark}
\subsection{Debiased log-likelihood function}\label{subsec:debiased}
When analyzing model~\eqref{eq:sampling}, the main challenge in estimating $\theta$ arises from the zeros generated in the unsuitable domain $\R^d\setminus S^*$. 
Since $Y>0$ implies $X\in S^*$, a natural approach is to use only the observations with $Y>0$. However, the corresponding likelihood function is not proper due to the selection. Let the quasi-likelihood function be
\begin{equation*}
	\ell(\theta;X,Y) = Y\log\sigma(X^\top\theta) +  (1-Y)\log(1-\sigma(X^\top\theta)).
\end{equation*}
The conditional expectation fails to satisfy $\theta^*=\argmax_{\theta}\E(\ell(\theta;X,Y)\mid Y>0)$, which motivates us to propose the debiased log-likelihood function. Proposition~\ref{prop:debiased} gives a general way to remove this bias.
\begin{proposition}\label{prop:debiased}
	Let $\cB$ be a Borel set on $\R$. If there exists a function $\phi(\theta;X)$ such that
	\begin{equation*}
		\nabla\phi(\theta;X)\big|_{\theta=\theta^*}
		=
		\E\big[\nabla\ell(\theta;X,Y)\mid Y\in\cB,X\big]\big|_{\theta=\theta^*},
		\qquad \forall \theta^*\in\Theta,
	\end{equation*}
	then $\theta^*$ is a stationary point of  $\E[\ell(\theta;X,Y)-\phi(\theta;X)\mid Y\in\cB,X]$.
\end{proposition}
\noindent
In this framework, we take the measurable set to be $\cB:=\R^+$. From Proposition~\ref{prop:debiased}, we have $\phi(\theta;X):=\log(1-(1-\sigma(X^\top\theta))^m)/m$. Thus, we propose the debiased log-likelihood function
\begin{equation}\label{eq:debiased_likelihood}
	\begin{split}
	\ell_{\con}(\theta;X,Y,m)
	={}&Y\log\sigma(X^\top\theta)+(1-Y)\log(1-\sigma(X^\top\theta))\\
	&-\frac{1}{m}\log\bigl(1-(1-\sigma(X^\top\theta))^m\bigr),
	\end{split}
\end{equation}
which leads to the Proposition~\ref{prop:de_function}.
\begin{proposition}\label{prop:de_function}
The debiased log-likelihood $\ell_{\con}(\theta;X,Y,m)$ is concave and has conditional expectation maximized at $\theta^*$.
\end{proposition}
Propositions~\ref{prop:debiased} and~\ref{prop:de_function} together give Fisher consistency. Concavity holds for every $(X,Y,m)$, so every stationary point of the population objective $\E[\ell_{\con}(\theta;X,Y,m)\mid Y>0]$ is a global maximizer, and Proposition~\ref{prop:debiased} places $\theta^*$ among the stationary points.
\begin{remark}
	$\ell_{\con}(\theta;X,Y,m)$ preserves a structure similar to that of the zero-truncated binomial distribution in \citet{johnson1993univariate}.
\end{remark}
\subsection{Reweighted support recovery}
\label{subsec:reweighted-support}
Next, we focus on recovering $S^*$. From Equation~\eqref{eq:sampling}, we have
\begin{equation}\label{eq:Z-law}
	\E[\I(Y>0)\mid X]=\bigl[1-(1-\sigma(X^\top\theta^*))^m\bigr]\,\I_{S^*}(X).
\end{equation}
This inspires us to construct a reweighted variable
\begin{equation}\label{eq:Z-def}
	Z=\frac{\I(Y>0)}{1-(1-\sigma(X^\top\theta^*))^m},
\end{equation}
which has the following property.

\begin{proposition}\label{prop:Z}
	Under Model~\eqref{eq:sampling}, we have
	\begin{equation*}
		\E[Z\mid X]=\I_{S^*}(X).
	\end{equation*}
	Consequently, for any $\tau\in(0,1)$, we have $\{X:\E[Z\mid X]>\tau\}=S^*$.
\end{proposition}

\begin{remark}
	Since $\E[Z\mid X]$ takes the value $1$ on $S^*$ and $0$ outside, it jumps at the boundary $\partial S^*$. Therefore, any fixed threshold $\tau\in(0,1)$ separates the two sides in the population.
\end{remark}
\newpage
\section{Estimation}\label{sec:estimation}
Let $\{(X_i,Y_i;m_i)\}_{i=1}^n$ be $n$ independent realizations from Model~\eqref{eq:sampling}. In this section, we present our approach for estimating the coefficient $\theta^*$ and the suitability region $S^*$ by $\hat{\theta}$ and $\hat{S}$, respectively, from the data $\{(X_i,Y_i;m_i)\}_{i=1}^n$. Roughly speaking, we use the positive observations to compute $\hat{\theta}$ via the debiased log-likelihood function~\eqref{eq:debiased_likelihood}. Then, using Proposition~\ref{prop:Z}, we construct a kernel estimator with the plug-in $\hat{\theta}$ to recover $\hat{S}$. The procedure is summarized in Algorithm~\ref{alg:main}.

\begin{algorithm}[h]
	\setstretch{1.5}
	\caption{Estimation procedure}\label{alg:main}
	\KwIn{$\{(X_i,Y_i;m_i)\}_{i=1}^n$, $\tau\in(0,1)$}
	$\hat{\theta}\leftarrow\arg\max_{\theta\in\Theta}L_{\con}(\theta)$\;
	$Z_i\leftarrow\dfrac{\I(Y_i>0)}{1-(1-\sigma(X_i^\top\hat{\theta}))^{m_i}}$\;
	$\hat{Z}(X)\leftarrow\dfrac{\sum_{i=1}^n K_h(X_i-X)Z_i}{\sum_{i=1}^n K_h(X_i-X)}$\;
	$\hat{S}\leftarrow\{X\in\R^d:\hat{Z}(X)>\tau\}$\;
	\KwOut{$\hat{\theta}$, $\hat{S}$}
\end{algorithm}
\subsection{Estimate the coefficient}\label{subsec:theta-est}

We first consider the regression component. Here $\ell_{\con}(\theta;X,Y)$ denotes the general debiased log-likelihood defined in \eqref{eq:debiased_likelihood}. Denote the indices of positive observations by
\begin{equation}\label{eq:pos_label}
	\cI:=\{i\in[n]:Y_i>0\},
\end{equation}
the sample objective function is defined as
\begin{equation}\label{eq:sample_loss}
	L_{\con}(\theta):=\frac{1}{M}\sum_{i\in\cI}m_i\ell_{\con}(\theta;X_i,Y_i,m_i),
\end{equation}
where $M:=\sum_{i\in\cI}m_i$. The coefficient is estimated by maximizing the objective function~\eqref{eq:sample_loss}, that is,
\begin{equation}\label{eq:theta-hat}
	\hat{\theta}:=\argmax_{\theta\in\Theta} L_{\con}(\theta).
\end{equation}
From Equation~\eqref{eq:debiased_likelihood}, $L_{\con}(\theta)$ is a sum of concave functions, which motivates the use of Newton's method to find the maximizer:
\begin{equation*}
	\theta^{(t+1)}
	=
	\theta^{(t)}
	-
	\left(\nabla^2 L_{\con}(\theta^{(t)})\right)^{-1}
	\nabla L_{\con}(\theta^{(t)}).
\end{equation*}
The optimization procedure is summarized in Algorithm~\ref{alg:theta}.

\begin{algorithm}[h]
	\setstretch{1.5}
	\caption{Optimization procedure}\label{alg:theta}
	\KwIn{$\nabla L_{\con}(\theta)$, $\nabla^2L_{\con}(\theta)$, $\epsilon>0$, $T\in\mathbb{N}$}
	Initialize $\theta^{(0)}\leftarrow\bm{0}$, $t\leftarrow 0$\;
	\Repeat{$\|\theta^{(t+1)}-\theta^{(t)}\|<\epsilon$ \textrm{or} $t=T$}{
		$\theta^{(t+1)}\leftarrow
		\theta^{(t)}
		-
		\left(\nabla^2L_{\con}(\theta^{(t)})\right)^{-1}
		\nabla L_{\con}(\theta^{(t)})$\;
		$t\leftarrow t+1$\;
	}
	\KwOut{$\hat{\theta}=\theta^{(t)}$}
\end{algorithm}

\subsection{Recovering the suitability region}\label{subsec:S-est}
After obtaining the estimate $\hat\theta$, we recover the suitability region via a plug-in approach. Specifically, we construct the reweighted outcome
\begin{equation}\label{eq:Z-est}
	Z_i=\frac{\I(Y_i>0)}{1-(1-\sigma(X_i^\top\hat\theta))^{m_i}},\qquad
	i=1,\dots,n,
\end{equation}
whose conditional mean equals $\I_{S^*}(X_i)$ when $\hat\theta=\theta^*$.
Then we estimate $\E[Z\mid X]$ by the Nadaraya--Watson estimator
\begin{equation}\label{eq:nw}
	\hat Z(X)=\frac{\sum_{i=1}^nK_h(X_i-X)\,Z_i}
	{\sum_{i=1}^nK_h(X_i-X)}.
\end{equation}
The support estimate is defined as $\hat S:=\{X:\hat Z(X)>\tau\}$, where $\tau\in(0,1)$ is a threshold for $\hat Z$.
By Proposition~\ref{prop:Z}, the population level of $\hat Z$ takes only the values $0$ and $1$, so every fixed $\tau\in(0,1)$ separates the two sides.

\newpage
\section{Theoretical results}\label{sec:theoretical}
In this section, we establish consistency results for $\hat\theta$ and $\hat{S}$. Benefiting from Proposition~\ref{prop:debiased}, we apply classical techniques to analyze the asymptotic behavior of $\hat\theta$. We then study the error bound for recovering $S$, which involves errors from both the standard kernel method and the plug-in approach.

\subsection{Asymptotic behavior of coefficient vector}\label{subsec:convergence_theta}

To begin with, we emphasize that the convergence analysis only depends on the positive outcomes $\{(X_i,Y_i;m_i)\}_{i\in\cI}$. Although $\cI$ is random, we condition on $\cI$ throughout this section and suppress this conditioning in the notation.

Intuitively, randomness enters through the positive outcomes, whose total effective sample size is $M:=\sum_{i\in\cI}m_i$. Hence, the expected convergence rate should be of order $M^{-1/2}$.
To establish the asymptotic results, we impose the following technical assumptions. Noting that the debiased log-likelihood function~\eqref{eq:debiased_likelihood} degenerates to zero when $m=1$, we impose Assumption~\ref{ass:sampling_time} to avoid trivial terms in Equation~\eqref{eq:sample_loss}.

\begin{assumption}\label{ass:sampling_time}
	The number of trials satisfies $\min_{i\in \cI}m_i\ge2$.
\end{assumption}
\begin{remark}
	In fact, the case $m_i=1$ does not hinder the estimation in the optimization process, but it contributes no new information to the objective function and only makes the convergence rate analysis more tedious.
\end{remark}
\noindent
Next, we examine the structure of the observed covariates. Equation~\eqref{eq:sample_loss} involves only the samples with positive outcomes. As a sample-level extension of Definition~\ref{def:ssd}, we impose Assumption~\ref{ass:design}.

\begin{assumption}\label{ass:design}
	There exists a constant $c\in\R^+$ such that the weighted covariance of $\{X_i\}_{i\in\cI}$ satisfies
	\begin{equation}\label{eq:bounded_hessian}
		\inf_{\theta\in\cB(\theta^*,r)}\lambda_{\mathrm{min}}\left[-\frac{1}{M}\sum_{i \in \cI} m_i\,\nabla^2\ell_{\con}(\theta;X_i,Y_i,m_i) \right]> c,
	\end{equation}
	where $\cB(\theta^*,r):=\{\theta\in\R^d:\|\theta-\theta^*\|_2\le r\}$ is a closed ball centered at $\theta^*$ with fixed radius $r$, and $\lambda_{\mathrm{min}}(\cdot)$ is the minimum eigenvalue of the matrix.
\end{assumption}

\noindent
To understand what Assumption~\ref{ass:design} captures, we compute the Hessian matrix of $\ell_{\con}(\theta;X,Y)$, which is
\begin{equation*}
	-\nabla^2\ell_{\con}(\theta;X,Y,m)
	=\frac{\sigma'(X^\top\theta)\Bigl[1-(1-\sigma(X^\top\theta))^m-m\sigma(X^\top\theta)(1-\sigma(X^\top\theta))^{m-1}\Bigr]}{\left[1-(1-\sigma(X^\top\theta))^m\right]^2}XX^\top.
\end{equation*}
Intuitively, Assumption~\ref{ass:design} requires a lower bound on the weighted average covariance matrix of the covariates $X$.
Since $\cI$ is the index set of positive outcomes, every $i\in\cI$ corresponds to $X_i\in S^*$. To obtain statistical guarantees, we require the covariance sequence $\{X_i\}_{i\in\cI}$ to be asymptotically non-degenerate, which is exactly what Assumption~\ref{ass:design} ensures.
\begin{remark}
	An alternative assumption such as $\E(XX^\top\mid S^*)\succ c\,I$ is not sufficient for consistency. A counterexample occurs when the trial counts $m_i$ are sufficiently unbalanced. Since the estimation should improve as $m_i$ increases, imposing an additional assumption such as bounded trial counts is counterintuitive and provides no further insight.
\end{remark}
\noindent
Furthermore, we constrain the scale of the covariate sequence $\{X_i\}_{i\in\cI}$ in Assumption~\ref{ass:covariate}.

\begin{assumption}\label{ass:covariate}
	The covariate vectors $\{X_i\}_{i\in\cI}$ satisfy
	\begin{equation}\label{eq:bounded_cov}
		\frac{1}{M}\sum_{i\in\cI}m_i\|X_i\|_2^2\le B.
	\end{equation} 
\end{assumption}

\noindent
This is a standard condition that controls the tail behavior of the covariates and is used to obtain uniform bounds on the estimator.
\begin{remark}
	In standard analyses, Assumption~\ref{ass:design} and Assumption~\ref{ass:covariate} are often stated in population form, for example, through the Fisher information matrix and the covariance matrix, which yields the usual $|\cI|^{-1/2}$ rate under a boundedness condition such as $\sup_{i\in\cI}m_i<\infty$. To obtain the expected $M^{-1/2}$ rate, we need weighted versions of these assumptions. To avoid introducing randomness in $m_i$, we state them directly in sample versions.
\end{remark}
\begin{theorem}\label{thm:consistency}
	Under Assumptions~\ref{ass:sampling_time}--\ref{ass:covariate}, as $M\to\infty$, with probability tending to one, $L_{\con}$ has a unique finite maximizer $\hat{\theta}$, and
	\begin{equation*}
		\|\hat{\theta} - \theta^*\|_2 = O_{\mathrm P}\left( M^{-1/2} \right).
	\end{equation*}
\end{theorem}
Next, we consider the asymptotic normality of $\hat\theta$. While consistency ensures that $\hat\theta$ converges to $\theta^*$, it does not characterize the fluctuations of $\hat\theta$ around $\theta^*$ at the $M^{-1/2}$ scale. To establish asymptotic normality, we need a uniform law of large numbers for the Hessian matrix in a neighborhood of $\theta^*$, together with a central limit theorem for the score function. These two requirements are formalized in Assumptions~\ref{ass:clt1} and~\ref{ass:clt2}, respectively.

\begin{assumption}\label{ass:clt1}
	There exists a positive definite matrix-valued function $J(\theta)$, continuous in $\theta$, such that
	\begin{equation*}
		\sup_{\theta\in\cB(\theta^*,r)}\left\|-\frac{1}{M}\sum_{i\in\cI} m_i\,\nabla^2\ell_{\con}(\theta;X_i,Y_i,m_i)-J(\theta)\right\|_F\to_{\mathrm P}0,
	\end{equation*}
	where $\cB(\theta^*,r)$ is the ball in Assumption~\ref{ass:design}.
\end{assumption}

\begin{assumption}\label{ass:clt2}
	Let $\xi_i=m_i\nabla\ell_{\con}(\theta^*;X_i,Y_i,m_i)$. For every $\varepsilon>0$,
	\[
	\frac1M\sum_{i\in\cI}
	\E\left[\|\xi_i\|_2^2
	\I\{\|\xi_i\|_2>\varepsilon\sqrt M\}\mid\mathcal F_n\right]
	\to_{\mathrm P}0,
	\]
	where $\mathcal F_n$ is the $\sigma$-field generated by $\cI$ and $\{(X_i,m_i):i\in\cI\}$.
\end{assumption}

\begin{remark}
	In fact, Assumptions~\ref{ass:clt1} and~\ref{ass:clt2} mainly constrain the behavior of $X_i$. Assumption~\ref{ass:clt1} only concerns the Hessian matrix, which is entirely unrelated to $Y_i$. Assumption~\ref{ass:clt2} is the standard Lindeberg condition for the weighted score.
\end{remark}
\begin{theorem}[Asymptotic normality]\label{thm:asymptotic_normality}
     Under the conditions of Theorem~\ref{thm:consistency} and Assumptions~\ref{ass:clt1}--\ref{ass:clt2},
	\[
		\sqrt{M}(\hat{\theta}-\theta^*)\ \to_{\mathrm d}\ \cN(\bm{0},J^{-1}(\theta^*)).
	\]
\end{theorem}
\subsection{Consistency of the support estimate}\label{subsec:support_consistency}

We now turn to the support estimator $\hat S=\{X:\hat Z(X)>\tau\}$, where $\hat Z$ is the Nadaraya--Watson estimator~\eqref{eq:nw} of the reweighted variable at the plug-in $\hat\theta$. Theorem~\ref{thm:consistency} requires no assumption on the distribution of $X$, because $\hat\theta$ pools all positive observations. The support estimator is local. The value $\hat Z(X)$ averages the reweighted outcomes of observations with $X_i$ near $X$, so it needs the $X_i$ to accumulate around $X$ as the bandwidth shrinks. We therefore assume that the distribution of $X$ has a density $f$, and measure the recovery error as follows.

\begin{definition}\label{def:support_error}
	The support recovery error of a set $S\subseteq\R^d$ is
	\[
	\mathcal{E}(S):=\int_{S\triangle S^*}f(X)\,dX.
	\]
\end{definition}

\noindent
The error $\mathcal{E}(S)$ is the probability that a new covariate falls where $S$ and $S^*$ disagree. It ignores regions without data and stays finite when $S^*$ is unbounded, as in Case~2 of Section~\ref{subsec:simulation}.

\begin{assumption}\label{ass:support_regular}
	The distribution of $X$ has a Lebesgue density $f$, and $\E\|X\|_2^2<\infty$.
\end{assumption}

\noindent
Neither boundedness, continuity nor compact support of $f$ is required, so Gaussian covariates are allowed. The moment bound is needed only for the rate in Theorem~\ref{thm:support}(ii).

For $X\in\R^d$ and $h,s\in(0,1]$, let $\cB(X,h)=\{X'\in\R^d:\|X'-X\|_2\le h\}$ and
\begin{equation*}
	\partial_hS^*:=\bigl\{X\in\R^d:\cB(X,h)\cap\partial S^*\neq\emptyset\bigr\},\quad
	L_{h,s}:=\Bigl\{X\in\R^d:\sigma(X^\top\theta^*)\int_{\cB(X,h)}f(X')\,dX'<sh^d\Bigr\}.
\end{equation*}
The set $\partial_hS^*$ is the layer of width $h$ around the boundary of $S^*$. $L_{h,s}$ is a region where few positive observations are expected.

\begin{assumption}\label{ass:support_rate}
	There exist constants $C,\alpha>0$ such that, for all $h,s\in(0,1]$,
	\begin{enumerate}
		\item[(a)] $\displaystyle\int_{\partial_hS^*}f(X)\,dX\le Ch$,
		\item[(b)] $\displaystyle\int_{L_{h,s}}f(X)\,dX\le C\bigl(s^\alpha+h\bigr)$.
	\end{enumerate}
\end{assumption}

\noindent
Assumption~\ref{ass:support_rate} controls the two regions where errors can arise. Windows centered in $\partial_hS^*$ mix observations from both sides of the boundary, and Assumption~\ref{ass:support_rate}(a) requires the probability of this layer to be of order $h$, as for a $(d-1)$-dimensional boundary. Assumption~\ref{ass:support_rate}(b) requires covariates to fall rarely in $L_{h,s}$, which holds when $f$ decays at least as fast as $\sigma(X^\top\theta^*)$. 

Next, we require the kernel in~\eqref{eq:nw} to be compactly supported and bounded away from zero near the origin.
\begin{assumption}\label{ass:kernel}
	There exist constants $0<\kappa\le\bar\kappa<\infty$ such that the kernel $K:\R^d\to[0,\infty)$ satisfies the following.
	\begin{enumerate}
		\item[(a)] $\displaystyle\int_{\R^d}K(u)\,du=1$, and $K(u)=0$ whenever $\|u\|_2>1$.
		\item[(b)] $K(u)\le\bar\kappa$ for all $u\in\R^d$, and $K(u)\ge\kappa$ for all $u\in\cB(\bm0,1/2)$.
	\end{enumerate}
\end{assumption}
\noindent
Therefore, we have the Theorem~\ref{thm:support} to capture the asymptotic behaviors of $\cE(\hat{S})$.

\begin{theorem}\label{thm:support}
	Suppose Assumptions~\ref{ass:support_regular} and~\ref{ass:kernel} hold, let $\tau\in(0,1)$ be fixed, and let $h_n$ tend to zero.
	\begin{enumerate}
		\item[(i)] If $\hat\theta$ converges to $\theta^*$ in probability and $nh_n^d$ tends to infinity, then
		\[
		\mathcal{E}(\hat S)\to_{\mathrm P}0.
		\]
		\item[(ii)] If, in addition, Assumption~\ref{ass:support_rate} holds, $\|\hat\theta-\theta^*\|_2=O_{\mathrm P}(n^{-1/2})$ and $nh_n^d/\log(1/h_n)$ tends to infinity, then
		\[
		\mathcal{E}(\hat S)=O_{\mathrm P}\biggl(h_n+\Bigl(\frac{\log(1/h_n)}{nh_n^d}\Bigr)^{\alpha}\biggr).
		\]
	\end{enumerate}
	Furthermore,	we obtain
	\[
	\mathcal{E}(\hat S)
	=O_{\mathrm P}\!\left(
	\left(\frac{\log n}{n}\right)^{\frac{\alpha}{\alpha d+1}}
	\right) \quad\text{with }h_n \asymp \left(\frac{\log n}{n}\right)^{\frac{\alpha}{\alpha d+1}}.
	\]
\end{theorem}

\begin{remark}
	Every trial carries information about $\theta^*$, so the rate of $\hat\theta$ is controlled by $M$. However, each observation contributes only one covariate location regardless of $m_i$, so the resolution of $\hat S$ is governed by $n$, and a larger $m_i$ only reduces the noise of $Z_i$. Since $M\ge|\cI|$ and $\P(Y>0)>0$ implies $|\cI|$ grows linearly in $n$, the condition $\|\hat\theta-\theta^*\|_2=O_{\mathrm P}(n^{-1/2})$ in Theorem~\ref{thm:support}(ii) follows from Theorem~\ref{thm:consistency}.
\end{remark}
\newpage
\section{Numerical study}\label{sec:numerical}

\subsection{Simulation}\label{subsec:simulation}

We verify the theoretical results in Section~\ref{sec:theoretical} with synthetic data. The covariate dimension and distribution are set to $d=2$ and $X\sim N(0,I_2)$, with $\theta^*=(1,1)$. Three different supports in Table~\ref{tab:cases} are tested in our study. All trials share a common size, $m_i\equiv m$. Data are generated from Model~\eqref{eq:sampling} with $m\in\{5,10,15\}$ and $n\in\{200,400,600,800,1000\}$, and each configuration is replicated 300 times.

\begin{table}[h]
	\centering
	\begin{tabular}{lll}
		\toprule
		Case & Support & Description \\
		\midrule
		1 & $S^*=\{X:\|X\|_2\le1\}$ & bounded \\
		2 & $S^*=\{X:\|X\|_2\ge1\}$ & unbounded \\
		3 & $S^*=\R^2$ & no structural zeros \\
		\bottomrule
	\end{tabular}
	\caption{Supports in the simulation.}
	\label{tab:cases}
\end{table}

\noindent\textbf{\underline{Estimation of $\theta^*$.}}
We compare four estimators. The \emph{oracle} is the MLE on the observations inside $S^*$. The \emph{naive} estimator is the MLE on all observations and ignores $S^*$. The \emph{uncorrected} estimator is the MLE on the positive observations without the truncation correction, which is an ablation of our method. Our estimator is the debiased MLE~\eqref{eq:theta-hat}.

\begin{table}[h]
	\centering
	{\setlength{\tabcolsep}{3pt}%
		\begin{tabular}{lccccccccc}
			\hline
			& \multicolumn{3}{c}{Case 1} & \multicolumn{3}{c}{Case 2} & \multicolumn{3}{c}{Case 3} \\
			\cmidrule(lr){2-4}\cmidrule(lr){5-7}\cmidrule(lr){8-10}
			$m$ & 5 & 10 & 15 & 5 & 10 & 15 & 5 & 10 & 15 \\
			\hline
			oracle      & .0228 & .0096 & .0075 & .0035 & .0017 & .0014 & .0031 & .0014 & .0010 \\
			naive       & 1.6681 & 1.6745 & 1.6921 & .0716 & .0729 & .0728 & \textbf{.0031} & \textbf{.0014} & \textbf{.0010} \\
			uncorrected & \underline{.0614} & \underline{.0106} & \textbf{.0074} & \underline{.0606} & \underline{.0134} & \underline{.0046} & .0582 & .0108 & .0032 \\
			ours        & \textbf{.0303} & \textbf{.0100} & \underline{.0076} & \textbf{.0056} & \textbf{.0020} & \textbf{.0015} & \underline{.0047} & \underline{.0017} & \underline{.0012} \\
			\hline
	\end{tabular}}
	\caption{Mean of $\|\hat\theta-\theta^*\|_2^2$ at $n=1000$ over 300 replications. Bold marks the best non-oracle estimator and underline the second best.}
	\label{tab:main}
\end{table}

Table~\ref{tab:main} reports the error at $n=1000$. Our estimator is best in most configurations. The naive estimator treats structural zeros as sampling zeros, so its bias does not vanish in Cases~1 and~2, while in Case~3 there are no structural zeros and it coincides with the oracle. As for the uncorrected estimator, since $\sigma(X^\top\theta^*)$ stays bounded away from zero on the bounded $S^*$, the debiasing weights approach one as $m$ grows, and in Case~1 it attains a rate similar to ours. In Case~2, however, $\sigma(X^\top\theta^*)$ can be arbitrarily close to zero, so the truncation bias persists and our estimator is best at every $m$. 

\begin{figure}[h]
	\centering
	\includegraphics[width=.8\textwidth]{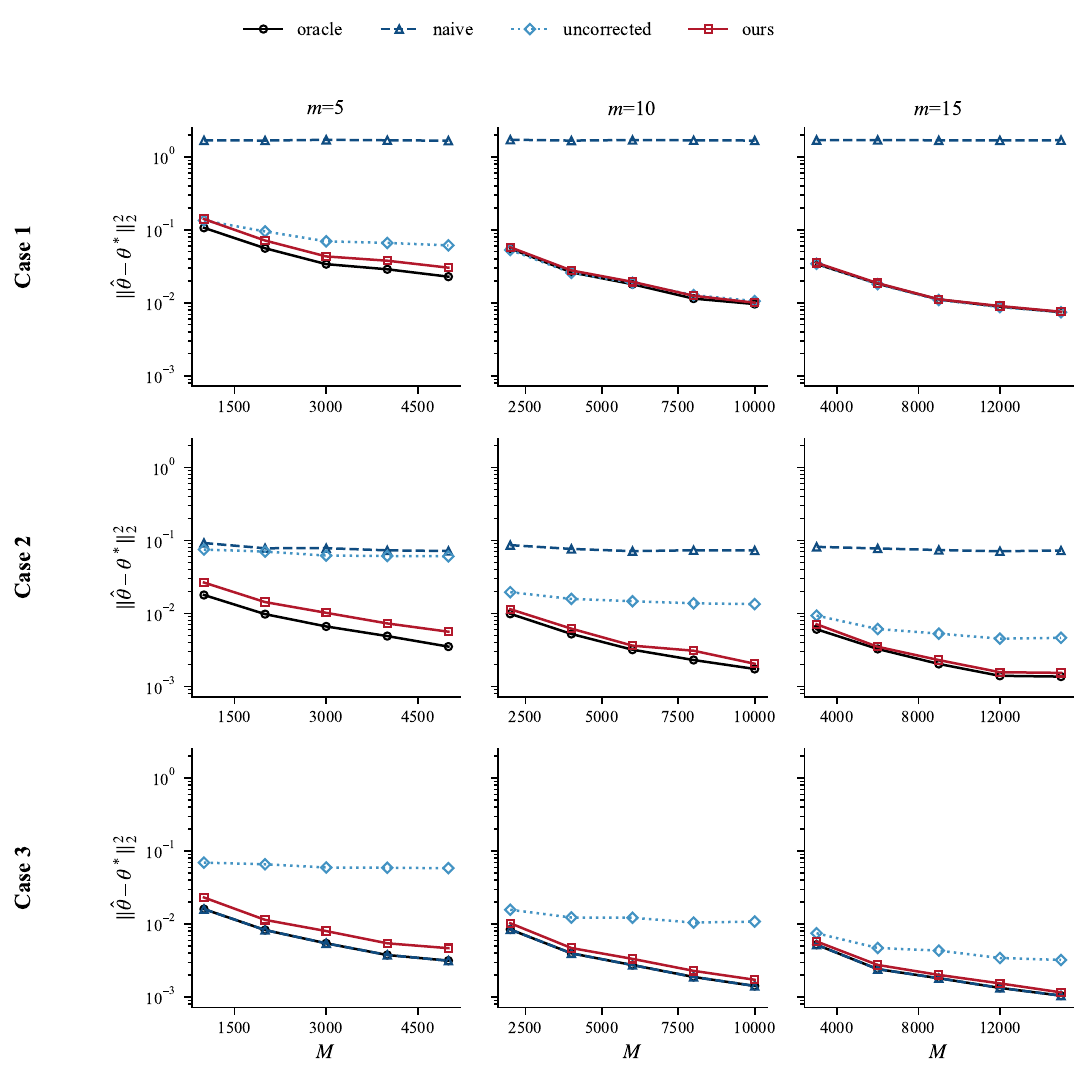}
	\caption{Mean of $\|\hat\theta-\theta^*\|_2^2$ versus $M=nm$. Rows are Cases 1--3 and columns are $m\in\{5,10,15\}$.}
	\label{fig:sim-theta}
\end{figure}

Figure~\ref{fig:sim-theta} shows how this error decays with the total number of trials $M=nm$, the effective sample size in Theorem~\ref{thm:consistency}.

\noindent\textbf{\underline{Estimation of the support.}}
To isolate the effect of the reweighting, we compare the reweighted outcome~\eqref{eq:Z-est} with the observed positivity indicator $\I(Y_i>0)$. Both approaches use the same kernel function, with bandwidth $h_n=(\log n/n)^{1/3}$ and threshold $\tau=1/2$. The simulation is designed to verify the theory and compare the approach with the vanilla method.

\begin{table}[h]
	\centering
	{\setlength{\tabcolsep}{5pt}%
		\begin{tabular}{lccccccccc}
			\hline
			& \multicolumn{3}{c}{Case 1} & \multicolumn{3}{c}{Case 2} & \multicolumn{3}{c}{Case 3} \\
			\cmidrule(lr){2-4}\cmidrule(lr){5-7}\cmidrule(lr){8-10}
			$m$ & 5 & 10 & 15 & 5 & 10 & 15 & 5 & 10 & 15 \\
			\hline
			uncorrected & .019 & \textbf{.014} & \textbf{.014} & .125 & .049 & .032 & .081 & .030 & .017 \\
			ours        & \textbf{.015} & \textbf{.014} & \textbf{.014} & \textbf{.037} & \textbf{.027} & \textbf{.022} & \textbf{.020} & \textbf{.012} & \textbf{.008} \\
			\hline
	\end{tabular}}
	\caption{Mean of $\mathcal{E}(\hat S)$ at $n=1000$ over 300 replications. Bold marks the smaller error, and ties are bold for both.}
	\label{tab:support}
\end{table}

Table~\ref{tab:support} reports the support results at $n=1000$. Overall, the behavior of the support estimate is similar to that of the coefficient $\hat\theta$. In Case~1, the two outcomes are nearly identical, whereas the observed positivity indicator $\I(Y_i>0)$ is outperformed in the other settings. 

\begin{figure}[h]
	\centering
	\includegraphics[width=.8\textwidth]{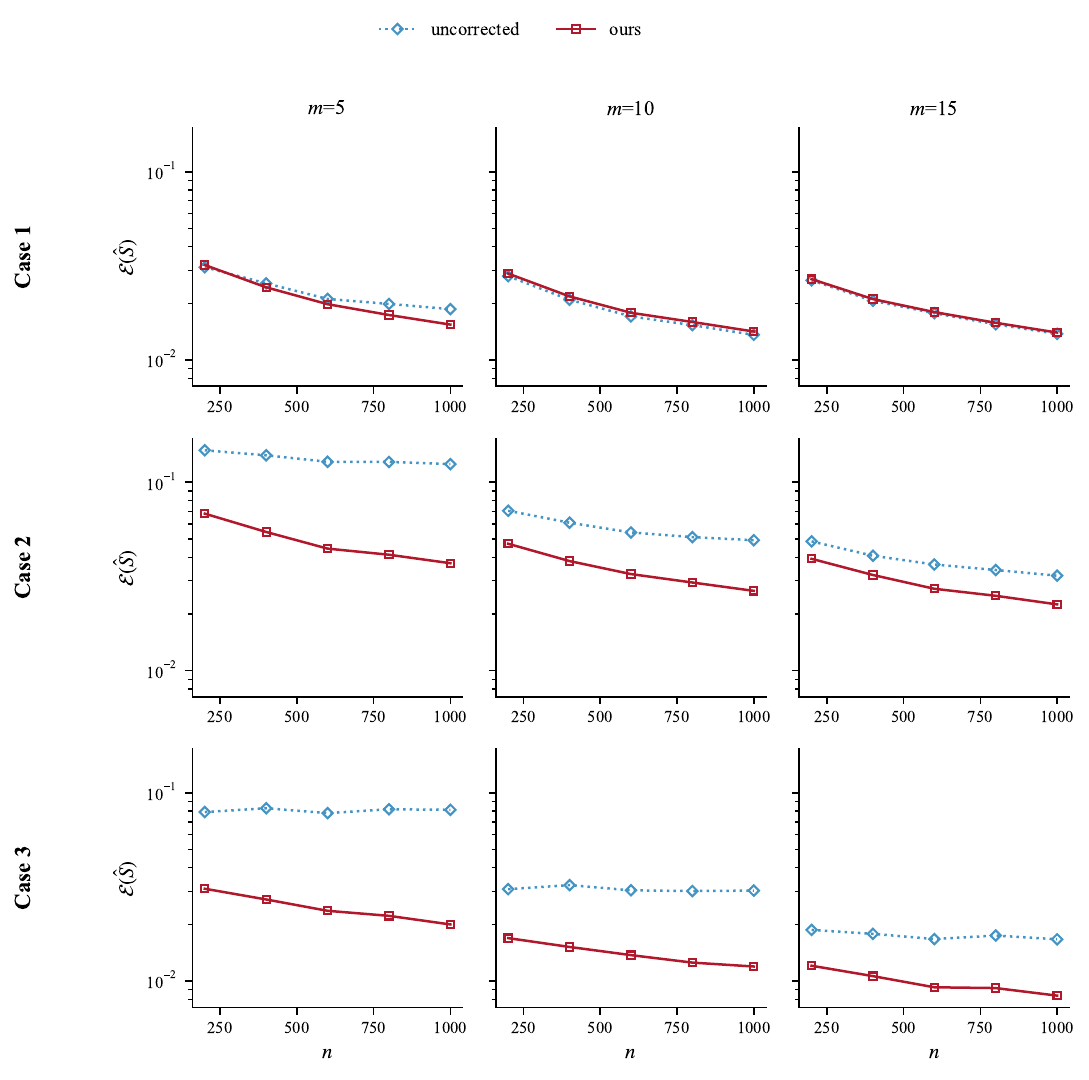}
	\caption{Mean of $\mathcal{E}(\hat S)$ versus $n$. Rows are Cases 1--3 and columns are $m\in\{5,10,15\}$.}
	\label{fig:sim-support-conv}
\end{figure}

Details of the convergence rate are shown in Figure~\ref{fig:sim-support-conv}.

\subsection{Real data analysis}\label{subsec:real-data}

In real data analysis, we consider the two datasets: \emph{trawl survey} and \emph{turtle nests}. In the trawl survey, structural zeros occur outside the thermal habitat and sampling zeros inside it, so the dataset exercises the full procedure. The turtle nests present the complementary regime: the sample is small  and the boundary is nearly linear in the covariates, so a nonlinear boundary offers little advantage and the kernel smoother operates with few observations.

\noindent\textbf{Methods.}
We use five cross-validation folds, fitting each method on the training part and evaluating it on the held-out part. All methods share the same folds, the same standardized covariates, and the same evaluation code, and every tuning choice is made inside the training fold. The kernel bandwidth $h$ is selected by leave-one-out cross-validation on the training set, and the threshold $\tau$ for the region $\hat{S}:=\{X\in\R^d:\hat{Z}(X)\ge\tau\}$ is fixed at $\tau=1/2$, as in the simulation. Covariates are standardized within the training fold, and real-data fits include an intercept via a constant column.

\noindent\textbf{Baselines.}
The competing methods are the two-part hurdle model \citep{cragg1971some,mullahy1986specification}, the zero-inflated binomial regression (ZIB) of \citet{hall2000zero}, and the zero-one inflated beta regression (ZOIB) of \citet{ospina2012general}, fitted with the R packages \texttt{glmmTMB} and \texttt{gamlss}. We also include TabICLv2 \citep{qu2026tabiclv2}, one of the most powerful foundation models for tabular data, through its reference implementation (package \texttt{tabicl}, version 2.0.2).

\noindent\textbf{Prediction criteria.}
We evaluate two tasks. The regression task compares the mean prediction $\hat Y$, scored by the unweighted mean absolute error (MAE). The classification task compares the positivity probability $\P(Y>0\mid X)$, scored by the area under the receiver operating characteristic curve (AUC) and the Matthews correlation coefficient (MCC), with each observation weighted by its $m_i$ trials and the score thresholded at the fixed level $1/2$. For the proposed method, we use
\begin{equation*}
	\hat Y_i:=\sigma(X_i^\top\hat\theta)\I_{\hat S}(X_i),\qquad \P(Y>0\mid X):=\left[1-(1-\sigma(X_i^\top\hat\theta))^{m_i}\right]\hat Z(X_i).
\end{equation*}
The baselines use their fitted mean and fitted positivity probability for both tasks.

\subsubsection*{\underline{Northeast US bottom trawl}}

The Northeast US bottom-trawl survey \citep{flanagan2018response} records hauls of the NMFS biannual survey from 1963 to 2015. After removing hauls with missing bottom temperature or depth, $n=30{,}131$ hauls remain, and 73 percent of them contain no cod. The outcome is the share of cod in the catch by numbers, $m_i$ is the total catch in numbers, and the covariates are bottom temperature ($0$ to $29^{\circ}$C) and depth ($5$ to $542$\,m).

\begin{table}[h]
	\centering
	\small
	\setlength{\tabcolsep}{5pt}
	\begin{tabular}{lccc}
		\toprule
		& MAE & AUC & MCC \\
		\midrule
		ours              & \textbf{.0167} (.000) & \textbf{.842} (.006) & \textbf{.375} (.021) \\
		hurdle            & .0236 (.000) & \underline{.493} (.012) & \underline{.004} (.017) \\
		ZIB               & \underline{.0232} (.000) & .492 (.013) & -.018 (.019) \\
		ZOIB              & .0298 (.000) & \underline{.493} (.012) & \underline{.004} (.017) \\
		TabICLv2          & .0266 (.004) & .492 (.006) & -.020 (.010) \\
		\bottomrule
	\end{tabular}
	\caption{Trawl ($n=30{,}131$), five-fold cross-validation. Bold marks the best method and underline the second best. Values in parentheses are standard errors over the five folds.}
	\label{tab:real-trawl}
\end{table}

We compare our method with the baselines. The five-fold cross-validation results are reported in Table~\ref{tab:real-trawl}. Our method attains the best MAE, ahead of all competitors. In positivity prediction, our score clearly dominates. 

To characterize the behavior of cod more directly, we fit the model to the full dataset. On the standardized scale, the fitted coefficients are $\hat\theta=(-0.56,-1.62,-4.13)$, corresponding to temperature, depth, and the intercept. Figure~\ref{fig:trawl-intro} reveals a sharp structural boundary at the margin, where the fitted suitability drops to zero beyond the observed thermal threshold.

The temperature results offer twofold support. The fitted coefficient is negative, and the estimated suitable region does not extend beyond \(18^{\circ}\)C. Both align with the documented thermal tolerance limit of Atlantic cod \citep{righton2010thermal}, confirming that the warm-water zeros are structural.

Rather than an intrinsic penalty of depth, the negative depth coefficient likely reflects a more complex ecological pattern. Figure~\ref{fig:trawl-intro} shows that suitability does not decline monotonically with depth, but instead forms a broad, curved band in cold waters. This pattern is consistent with the seasonal \emph{vertical migration} of Atlantic cod, which move to deeper waters to overwinter and return to shallower feeding grounds in summer \citep{lawson2000seasonal,neat2006residency}. Because the biannual survey pools winter and summer hauls, the negative depth coefficient likely absorbs this seasonal contrast---deep winter hauls versus shallow summer hauls---rather than reflecting a direct effect of depth alone.

\subsubsection*{\underline{Loggerhead sea turtle nests}}

The loggerhead data \citep{carroll2022loggerhead} record 170 nests on Ossabaw Island, Georgia, of which 164 have complete covariates and 28 percent failed to hatch. The outcome is the hatching proportion, $m$ is the number of eggs (47 to 177 per nest), and the covariates are nest temperature and elevation.

\begin{table}[h]
	\centering
	\small
	\setlength{\tabcolsep}{5pt}
	\begin{tabular}{lccc}
		\toprule
		& MAE & AUC & MCC \\
		\midrule
		ours              & \textbf{.2116} (.006) & \underline{.837} (.036) & \underline{.456} (.094) \\
		hurdle            & .2232 (.004) & .834 (.031) & .375 (.082) \\
		ZIB               & .2232 (.004) & .834 (.031) & .375 (.082) \\
		ZOIB              & .2258 (.004) & .834 (.031) & .375 (.082) \\
		TabICLv2          & \underline{.2166} (.006) & \textbf{.845} (.038) & \textbf{.472} (.106) \\
		\bottomrule
	\end{tabular}
	\caption{Loggerhead nests ($n=164$), five-fold cross-validation. Bold marks the best method and underline the second best. Values in parentheses are standard errors over the five folds.}
	\label{tab:real-turtle}
\end{table}

Table~\ref{tab:real-turtle} reports the five-fold cross-validation results. Our method remains competitive, attaining the best MAE, with AUC and MCC close to TabICLv2 while outperforming the parametric baselines. The margin is smaller than in the trawl survey primarily because the sample is drastically smaller (170 nests versus 30{,}131 hauls), which limits the advantage of a complex nonlinear boundary. TabICLv2 edges ahead on MCC, which is expected given its design as a foundation model tailored for small datasets and its large parameter count.

The fitted coefficients are $\hat\theta=(0.44,0.28,0.22)$, corresponding to temperature, elevation, and the intercept, all positive within $\hat S$. Temperature is positively associated with incubation, with warmer nests hatching at higher rates within the observed thermal range \citep{godley2001thermal}. Elevation is associated with exposure, as higher nests are shielded from the tidal inundation and erosion that cause total failure \citep{whitesell2022direct}.

\begin{figure}[h]
	\centering
	\includegraphics[width=\textwidth]{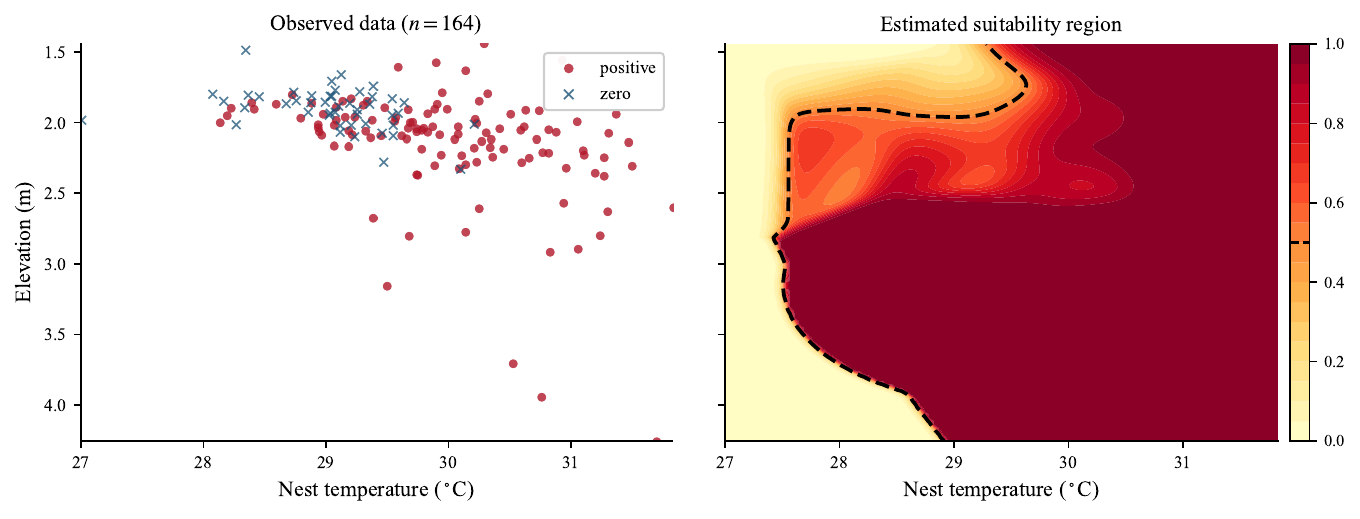}
	\caption{Loggerhead nests. Left: observed data. Right: the suitable region recovered by our method.}
	\label{fig:turtle-support}
\end{figure}

Unfortunately, the boundary of the estimated region \(S\) is somewhat diffuse, with several positive samples scattered outside. 
This limitation stems partly from the small sample size, which restricts the precise recovery of a strict boundary. 
More fundamentally, this may reflect the ecological reality of the system. 
Nest failure may be driven mainly by tidal washovers and wave disturbance, which strike nests largely regardless of temperature and elevation \citep{whitesell2022direct,ceriani2026wave}.
Because wave-driven flooding occurs stochastically across all zones \citep{ceriani2026wave}, it precludes the formation of any fixed unsuitable region.

\medskip
\noindent\textbf{Summary.}
Across both datasets, our simple model nearly matches or outperforms every competing method, including the tabular foundation model and the parametric baselines. Beyond prediction, the estimated unsuitable region explains the behavior of zero outcomes and enhances our understanding of the data.

\newpage
\section{Discussion}\label{sec:discussion}

This paper proposes a hybrid model for zero-inflated proportion data and treats the structural-zero component as a deterministic covariate region rather than a smooth mixture.  In most statistical models, uncertainty is attributed to randomness. However, in many applications in physics, a substantial part of the phenomenon determined by a partial differential equation is entirely non-randomly. Thus, extending the traditional randomness-based perspective to incorporate deterministic modeling is necessary. 

However, some issues remain worth thinking about. First, from the turtle data, we find that recovering $S$ is not straightforward. The recovered region depends on how the bandwidth and the threshold are chosen, and resolving the boundary also requires a sufficiently rich set of covariate locations. Second, the method combines a generalized linear model with a kernel smoother, and each has its own limitation. The former forces the response to be monotone along a single direction, while the latter is practical only in low dimensions, so neither can capture a boundary that varies with several covariates in a complex way.

From a longer horizon, the issues raised above are less consequential than they first appear. In the era of flourishing neural networks, many highly promising models are now available, while data-driven questions remain. Using these models interpretably from a statistical perspective, and solving real-world problems with them, appears all the more important.

\section*{Data and code availability}

The trawl data of Section~\ref{subsec:real-data} are publicly
available from the Dryad repository
(\texttt{doi:}\allowbreak\texttt{10.5061/}\allowbreak\texttt{dryad.}\allowbreak\texttt{bg08269}), and
the loggerhead nest data from the Georgia Southern University
Digital Commons. The code reproducing all tables and figures is
available at
\url{https://github.com/PinjunD/Hybrid-model-for-zero-inflated-proportion-data}.
\newpage

\bibliographystyle{plainnat}
\bibliography{refs}

\newpage
\appendix
\section{Proofs of Propositions}
\subsection{Proof of Proposition~\ref{prop:iden}}\label{proof_iden}

\begin{proof}
	Suppose there exist $(\theta,S)$ and $(\theta',S')$ satisfying
	\begin{equation}\label{eq:iden}
		\sigma(X^\top\theta)\I_S(X)
		=
		\sigma(X^\top\theta')\I_{S'}(X)
		\qquad \text{for all }X\in\R^d.
	\end{equation}
	We prove both directions: the full-span condition is sufficient for identifiability, and it is also necessary.
	
	\subsection*{Sufficiency}
	Take $X\in S\setminus S'$, which implies $\I_{S'}(X)=0$. From Equation~\eqref{eq:iden}, we have $\sigma(X^\top\theta)\I_S(X)=0$, which is impossible since $\sigma\in(0,1)$ and $X\in S$.
	Hence $S\setminus S'=\emptyset$, so $S\subseteq S'$. Similarly, we have $S'\subseteq S$, which implies $S=S'$.
	Next, we take $X\in S$. From Equation~\eqref{eq:iden}, we have $\sigma(X^\top\theta)=\sigma(X^\top\theta')$.
	Due to the monotonicity of $\sigma$, we obtain
	\begin{equation*}
		X^\top(\theta-\theta')=0, \quad\forall X\in S.
	\end{equation*}
	Hence, we have $\theta-\theta'\in\spa(S)^\perp$. Since $S$ is full-span, it follows that $\spa(S)^\perp=\{0\}$. Thus, we have $\theta=\theta'$.
	Therefore, the model is identifiable over $\Theta\times\cS$.
	
	\subsection*{Necessity}
	If $S$ is not full-span, then $\spa(S)^\perp\neq\{0\}$, so there exists a nonzero vector $v\in\R^d$ such that $X^\top v=0$ for any $X\in \spa(S)$. For any given $c\neq 0$, define $\theta':=\theta+c v\neq \theta$. We have
	\begin{equation}\label{eq:iden2}
		\sigma(X^\top\theta')\I_{S}(X) = \sigma(X^\top\theta)\I_{S}(X).
	\end{equation}
	When $X\in S$, we have $X^\top v=0$, and hence Equation~\eqref{eq:iden2} holds.
	When $X\notin S$, we have $\I_{S}(X)=0$, and hence Equation~\eqref{eq:iden2} also holds. Hence $(\theta,S)$ and $(\theta',S)$ induce the same model with $\theta'\neq\theta$, so identifiability fails.
\end{proof}

\subsection{Proof of Proposition~\ref{prop:debiased}}\label{proof_debiased}

\begin{proof}
	Let $\cB$ be a Borel set. Suppose $\phi(\theta;X)$ satisfies
	\begin{equation*}
		\nabla\phi(\theta;X)\big|_{\theta=\theta^*}
		=
		\E\big[\nabla\ell(\theta;X,Y)\mid Y\in\cB,X\big]\big|_{\theta=\theta^*},
		\qquad \forall \theta^*\in\Theta.
	\end{equation*}
	The debiased function $l(\theta;X,Y):=\ell(\theta;X,Y)-\phi(\theta;X)$ satisfies
	\begin{equation*}
		\nabla l(\theta;X,Y)=\nabla\ell(\theta;X,Y)-\nabla\phi(\theta;X),
	\end{equation*}
	which implies
	\begin{equation*}
		\E\big[\nabla l(\theta;X,Y)\mid Y\in\cB,X\big]=\E\big[\nabla\ell(\theta;X,Y)\mid Y\in\cB,X\big]-\nabla\phi(\theta;X).
	\end{equation*}
	Thus, we have $\E\big[\nabla l(\theta;X,Y)\mid Y\in\cB,X\big]\big|_{\theta=\theta^*}=\bm0$ for any $\theta^*\in\Theta$, which implies $\theta^*$ is the stationary point.
\end{proof}

\subsection{Proof of Proposition~\ref{prop:de_function}}\label{asec:cond}

\begin{proof}
	Let $\cB:=\R^+$. The conditional expectation satisfies
	\begin{align*}
		\E\big[\nabla\ell(\theta;X,Y)\mid Y\in\cB,X\big]&=\E\left[\nabla\left(Y\log\sigma(X^\top\theta)+(1-Y)\log(1-\sigma(X^\top\theta))\right)\Bigm| Y\in\cB,X\right]\\
		&=\E\left[\left(Y-\sigma(X^\top\theta)\right)X\Bigm| Y\in\cB,X\right]\\
		&=\left[\E(Y\mid Y\in\cB)-\sigma(X^\top\theta)\right]X\\
		&=\left[\frac{\sigma(X^\top\theta^*)}{1-(1-\sigma(X^\top\theta^*))^m}-\sigma(X^\top\theta)\right]X.
	\end{align*}
	Thus, we have
	\begin{align*}
		\E\big[\nabla\ell(\theta;X,Y)\mid Y\in\cB,X\big]\big|_{\theta=\theta^*}
		=& \frac{\sigma(X^\top\theta^*)(1-\sigma(X^\top\theta^*))^{m}}{1-(1-\sigma(X^\top\theta^*))^m}X\\
		=&\frac{\sigma'(X^\top\theta^*)(1-\sigma(X^\top\theta^*))^{m-1}}{1-(1-\sigma(X^\top\theta^*))^m}X\\
		=&\nabla\phi(\theta;X)\big|_{\theta=\theta^*},
	\end{align*}
	where $\phi(\theta;X)$ exactly satisfies
	\[
	\phi(\theta;X) := \frac{1}{m}\log(1-(1-\sigma(X^\top\theta))^m).
	\]
	Therefore, we have
	\[
	\ell_{\con}(\theta;X,Y,m) = Y\log\sigma(X^\top\theta)+(1-Y)\log(1-\sigma(X^\top\theta))-\frac{1}{m}\log(1-(1-\sigma(X^\top\theta))^m).
	\]
	The gradient and Hessian of $\ell_{\con}(\theta;X,Y)$ are
	\begin{equation}\label{eq:gradient}
		\nabla\ell_{\con}(\theta;X,Y,m)= \left[Y-\frac{\sigma(X^\top\theta)}{1-(1-\sigma(X^\top\theta))^m}\right]X,
	\end{equation}
	\begin{equation*}
		\begin{split}
		\nabla^2\ell_{\con}(\theta;X,Y,m)
		={}&{-}\frac{\sigma'(X^\top\theta)}{\left[1-(1-\sigma(X^\top\theta))^m\right]^2}\\
		&\times\Bigl[1-(1-\sigma(X^\top\theta))^m-m\sigma(X^\top\theta)(1-\sigma(X^\top\theta))^{m-1}\Bigr]XX^\top.
	\end{split}
	\end{equation*}
Since $(1-x)^m+mx(1-x)^{m-1}$ is monotonically decreasing in $m$ for $m\ge 2$,
we have
\begin{equation*}
	(1-x)^m+mx(1-x)^{m-1}\le (1-x)^2+2x(1-x)=1-x^2,
\end{equation*}
which implies
\begin{equation}\label{eq:concave}
	-\nabla^2\ell_{\con}(\theta;X,Y,m)
	\succeq\frac{\sigma'(X^\top\theta)\sigma^2(X^\top\theta)}{\left[1-(1-\sigma(X^\top\theta))^m\right]^2}\cdot XX^\top
	\succeq
	\sigma'(X^\top\theta)\sigma^2(X^\top\theta)\cdot XX^\top
    \succeq 0.
\end{equation}
Thus, $\ell_{\con}(\theta;X,Y,m)$ is a concave function.
	From Proposition~\ref{prop:debiased}, we have
	\[
	\E(\nabla\ell(\theta;X,Y,m)\mid Y>0,X)\big|_{\theta=\theta^*}=\bm0.
	\]
	By concavity, it follows that $\ell_{\con}(\theta;X,Y,m)$ is proper.\footnote{Rigorously, uniqueness of $\theta$ requires the full-rank structure of $X$, by a standard linear algebra argument.}
\end{proof}

\subsection{Proof of Proposition~\ref{prop:Z}}\label{asec:Z}

\begin{proof}
From Equation~\eqref{eq:sampling}, the conditional probability satisfies
\[\P(Y>0\mid X)=\bigl[1-(1-\sigma(X^\top\theta^*))^m\bigr]\I_{S^*}(X).\]
Hence, we have
\[
\E[Z\mid X]
=\frac{\P(Y>0\mid X)}{1-(1-\sigma(X^\top\theta^*))^m}
=\I_{S^*}(X).
\]
Note that $\I_{S^*}(X)\in\{0,1\}$, we have
\[\{X\in\R^d:\E[Z\mid X]>\tau\}=\{X\in\R^d:\I_{S^*}(X)>\tau\}=S^*\] for any $\tau\in(0,1)$.
\end{proof}

\section{Proof of theorems}\label{proof:thm}

\subsection{Proof of Theorem~\ref{thm:consistency}}\label{asec:consistency}

We first recap the notation. The observed data $\{(X_i,Y_i;m_i)\}_{i=1}^n$ follows Equation~\eqref{eq:sampling}. Let $\cI:=\{i\in[n]:Y_i>0\}$ and $M:=\sum_{i\in\cI}m_i$. The sample objective function in Equation~\eqref{eq:sample_loss} is
\begin{equation*}
	L_\con(\theta):=\frac1M\sum_{i\in\cI} m_i\ell_\con(\theta;X_i,Y_i,m_i).
\end{equation*}
We prove that, with probability tending to one, this objective admits a unique maximizer and that $\|\hat\theta-\theta^*\|_2=O_{\mathrm P}(M^{-1/2})$. Throughout the proof, $d$ is fixed and $M\to\infty$ in probability.

\medskip
\noindent\textbf{Step 1: Control the score.}

Let $\mathcal F_n$ denote the sigma-field generated by $\cI$ and $\{(X_i,m_i):i\in\cI\}$. We use $\bar L_{\con}(\theta):=\E(L_{\con}(\theta)\mid\mathcal F_n)$ to denote the corresponding conditional expectation.
Noting that
\begin{equation}\label{eq:con_Y}
	\E(Y_i\mid Y_i>0,X_i,m_i)=\frac{\sigma(X_i^\top\theta^*)}{1-(1-\sigma(X_i^\top\theta^*))^{m_i}},
\end{equation}
we have
\begin{align*}
	L_{\con}(\theta)&=\frac{1}{M}\sum_{i\in\cI}\left(m_iY_i\log\frac{\sigma(X_i^\top\theta)}{1-\sigma(X_i^\top\theta)}
	+\log\frac{(1-\sigma(X_i^\top\theta))^{m_i}}{1-(1-\sigma(X_i^\top\theta))^{m_i}}\right),\\
	\bar{L}_{\con}(\theta)&\stackrel{\eqref{eq:con_Y}}{=}\frac{1}{M}\sum_{i\in\cI}\left(\frac{m_i\sigma(X_i^\top\theta^*)}{1-(1-\sigma(X_i^\top\theta^*))^{m_i}}\log\frac{\sigma(X_i^\top\theta)}{1-\sigma(X_i^\top\theta)}
	+\log\frac{(1-\sigma(X_i^\top\theta))^{m_i}}{1-(1-\sigma(X_i^\top\theta))^{m_i}}\right).
\end{align*}
Denoting $u_i:=\E(Y_i\mid Y_i>0,X_i,m_i)$, the empirical process is given by
\begin{equation}\label{eq:centered_process}
	G(\theta):=L_{\con}(\theta)-\bar{L}_{\con}(\theta)
	=\frac1M\sum_{i\in\cI}m_i(Y_i-u_i)\log\frac{\sigma(X_i^\top\theta)}{1-\sigma(X_i^\top\theta)}.
\end{equation}
Since $\sigma(x)$ is the sigmoid function, Equation~\eqref{eq:centered_process} simplifies to
\begin{equation}\label{eq:linear_process}
	G(\theta)=\frac1M\sum_{i\in\cI}m_i(Y_i-u_i)X_i^\top\theta.
\end{equation}
For any $x\in(0,1)$ and $m\ge1$, we have
\begin{align*}
	&(1-x)\cdot\frac{1-(1-x)^{m}}{x}=(1-x)\sum^{m-1}_{k=0}(1-x)^k\le m\\
	\implies&(1-x)^{m}\left[1-(1-x)^{m}\right]\le mx(1-x)^{m-1}\\
	\implies& 1-(1-x)^{m}-mx(1-x)^{m-1}\le\left[1-(1-x)^{m}\right]^2.
\end{align*}
Let $p_i:=\sigma(X_i^\top\theta^*)$. Taking $x=p_i$ and $m=m_i$, we have
\begin{align*}
	\E((Y_i-u_i)^2\mid Y_i>0)
	&=\frac{p_i(1-p_i)}{m_i\left[1-(1-p_i)^{m_i}\right]}-\frac{p_i^2(1-p_i)^{m_i}}{\left[1-(1-p_i)^{m_i}\right]^2}\\
	&=\frac{p_i(1-p_i)\left[1-(1-p_i)^{m_i}-m_ip_i(1-p_i)^{m_i-1}\right]}{m_i\left[1-(1-p_i)^{m_i}\right]^2}\\
	&\le\frac{p_i(1-p_i)}{m_i}\le\frac{1}{4m_i}.
\end{align*}
From Assumption~\ref{ass:covariate}, we have
\begin{equation}\label{eq:markov}
	\P\left(
	\left\|\sum_{i\in\cI}m_i(Y_i-u_i)X_i\right\|_2^2
	\ge \frac B\delta \cdot M
	\,\middle|\,\mathcal F_n\right)
	\le
	\frac{\delta}{4B M}\sum_{i\in\cI}m_i\|X_i\|_2^2\stackrel{\eqref{eq:bounded_cov}}{\le} \delta.
\end{equation}
Let $\theta\in\cB(\theta^*,r):=\{\theta:\|\theta-\theta^*\|_2\le r\}$. Since $G(\theta) = \theta^\top\left[\sum_{i\in\cI}m_i(Y_i-u_i)X_i\right]/M$,
we have
\begin{align*}
	\sup_{\theta\in\cB(\theta^*,r)}
	|G(\theta)-G(\theta^*)|\le\frac{r}{M}
	\left\|\sum_{i\in\cI}
	m_i(Y_i-u_i)X_i\right\|_2.
\end{align*}
Thus, for any $\delta>0$, there exists a constant $c_1:= r\sqrt{B/\delta}$ such that
\begin{equation}\label{eq:bound1}
	\sup_{\theta\in\cB(\theta^*,r)}|G(\theta)-G(\theta^*)|\stackrel{\eqref{eq:markov}}{\le} \frac{c_1}{\sqrt M}\qquad\text{with probability at least $1-\delta$.}
\end{equation}
Notice that
\begin{equation*}
	\nabla^2\bar{L}_{\con}(\theta)=\nabla^2{L}_{\con}(\theta)=\frac1M\sum_{i\in\cI}m_i\nabla^2\ell_{\con}(\theta;X_i,Y_i,m_i).
\end{equation*}
According to Assumption~\ref{ass:design} and $\nabla\bar L_{\con}(\theta^*)=\bm0$ by \eqref{eq:con_Y}, $\bar{L}_{\con}(\theta)$ is strongly concave on $\cB(\theta^*,r)$ and maximized at $\theta^*$.
Thus,
\begin{equation*}
	\bar{L}_{\con}(\theta^*)-\bar{L}_{\con}(\theta)\ge \frac c2\|\theta-\theta^*\|^2_2.
\end{equation*}
Furthermore, with probability at least $1-\delta$, we have
\begin{align}\label{eq:bound2}
	{L}_{\con}(\theta^*)-{L}_{\con}(\theta)
	=\bar{L}_{\con}(\theta^*)-\bar{L}_{\con}(\theta)+{G}(\theta^*)-{G}(\theta)\stackrel{\eqref{eq:bound1}}{\ge}\frac{c}{2}\|\theta-\theta^*\|^2_2-\frac{c_1}{\sqrt M}.
\end{align}
Take $r^2_0:=2c_1/\sqrt{c^2M}\rightarrow0$. For any $\theta\in\cB(\theta^*,r)\setminus\cB(\theta^*,r_0)$, we have ${L}_{\con}(\theta^*)-{L}_{\con}(\theta)>0$ by Equation~\eqref{eq:bound2}, which implies $\hat\theta\notin\cB(\theta^*,r)\setminus\cB(\theta^*,r_0)$.
Furthermore, for any $\theta\in\Theta\setminus\cB(\theta^*,r)$, there exists $\theta_t\in\cB(\theta^*,r)\setminus\cB(\theta^*,r_0)$ such that $\theta_t:=t\theta+(1-t)\theta^*$.
From Equation~\eqref{eq:concave}, we know that
\begin{equation*}	-\nabla^2{L}_{\con}(\theta)\succeq\frac1M\sum_{i\in\cI}m_i\sigma'(X_i^\top\theta)\sigma^2(X_i^\top\theta)\cdot X_iX_i^\top
	\succeq 0,
\end{equation*}
which implies ${L}_{\con}(\theta)$ is globally concave.
If ${L}_{\con}(\theta^*)\le{L}_{\con}(\theta)$, based on the global concavity, we have
\begin{equation*}
	{L}_{\con}(\theta_t)\ge t{L}_{\con}(\theta)+(1-t){L}_{\con}(\theta^*)\ge {L}_{\con}(\theta^*),
\end{equation*}
which leads to a contradiction. Therefore, we have $\hat\theta\in\cB(\theta^*,r_0)$, namely,
\begin{equation}\label{eq:rough_bound}
	\|\hat\theta-\theta^*\|_2^2\le\frac{2c_1}{c}\cdot\frac{1}{\sqrt M}.
\end{equation}

\medskip
\noindent\textbf{Step 2: Enhance the convergence rate.}

We consider the gradient of $L_{\con}$ at $\theta^*$. By Equation~\eqref{eq:gradient} and~\eqref{eq:con_Y},
\begin{equation*}
	\nabla L_{\con}(\theta^*)=\frac{1}{M}\sum_{i\in\cI}m_i(Y_i-u_i)X_i.
\end{equation*}
Thus, for any $\delta>0$, there exists a constant $c_3:=\sqrt{B/\delta}$ such that, with probability $1-\delta$,
\begin{equation}\label{eq:convergence_gradient}
	\|\nabla L_{\con}(\theta)\|_2
	:=
	\left\|\frac{1}{M}\sum_{i\in\cI}m_i(Y_i-u_i)X_i\right\|_2
	\overset{\eqref{eq:markov}}{\le}
	\frac{c_3}{\sqrt M}.
\end{equation}
By integration, we have
\begin{equation}\label{eq:taylor}
	\nabla L_{\con}(\theta^*)+\left[\int_{0}^{1}\nabla^2L_{\con}(\theta^*+(\hat\theta-\theta^*)t)dt\right](\hat\theta-\theta^*)=\nabla L_{\con}(\hat\theta)=\bm 0.
\end{equation}
From Equation~\eqref{eq:rough_bound}, we have $\hat\theta\in\cB(\theta^*,r_0)\subset \cB(\theta^*,r)$. Thus, for any $t\in[0,1]$, $\theta^*+(\hat\theta-\theta^*)t\in \cB(\theta^*,r)$.
According to Assumption~\ref{ass:design}, we have
\begin{align*}
	\|\nabla L_{\con}(\theta^*)\|_2
	&=\left\|\left[\int_{0}^{1}-\nabla^2L_{\con}(\theta^*+(\hat\theta-\theta^*)t)dt\right](\hat\theta-\theta^*)\right\|_2\ge\lambda_{\mathrm{min}}\|\hat\theta-\theta^*\|_2\ge c\|\hat\theta-\theta^*\|_2.
\end{align*}
Here, $\lambda_{\mathrm{min}}$ is the minimum eigenvalue of $\int_{0}^{1}-\nabla^2L_{\con}(\theta^*+(\hat\theta-\theta^*)t)dt$.
From the asymptotic behavior of $\nabla L_{\con}(\theta)$ in \eqref{eq:convergence_gradient}, with probability $1-\delta$, we have
\begin{equation*}
	\|\hat\theta-\theta^*\|_2\le\frac{c_3}{c}\cdot\frac{1}{\sqrt M}.
\end{equation*}
This completes the proof.
\qed

\subsection{Proof of Theorem~\ref{thm:asymptotic_normality}}
Noting that
\begin{align*}
	\E((Y_i-u_i)^2\mid Y_i>0)=\frac{p_i(1-p_i)}{m_i\cdot\left[1-(1-p_i)^{m_i}\right]}-\frac{p^2_i(1-p_i)^{m_i}}{\left[1-(1-p_i)^{m_i}\right]^2},
\end{align*}
and that $\{Y_i\}_{i\in\cI}$ are conditionally independent given $\mathcal F_n$, the covariance matrix of $\nabla L_{\con}(\theta^*)$ is given by
\begin{equation*}
	\Cov[\nabla L_{\con}(\theta^*)\mid\mathcal F_n]
	=
	\frac1{M^2}\sum_{i\in\cI}\frac{m_ip_i(1-p_i)\left[1-(1-p_i)^{m_i}-m_ip_i(1-p_i)^{m_i-1}\right]}{\left[1-(1-p_i)^{m_i}\right]^2} X_iX_i^\top.
\end{equation*}
Next, the Hessian matrix is given by
\begin{equation*}
	\nabla^2L_{\con}(\theta)
	=
	\frac1M\sum_{i\in\cI}\frac{-m_i\sigma'(X_i^\top\theta)\left[1-(1-\sigma(X_i^\top\theta))^{m_i}-m_i\sigma(X_i^\top\theta)(1-\sigma(X_i^\top\theta))^{m_i-1}\right]}{\left[1-(1-\sigma(X_i^\top\theta))^{m_i}\right]^2} X_iX_i^\top.
\end{equation*}
Thus, we have $M\cdot\Cov[\nabla L_{\con}(\theta^*)\mid\mathcal F_n]=-\nabla^2L_{\con}(\theta^*)$.
From Equation~\eqref{eq:taylor}, we have
\begin{equation*}
	\hat\theta-\theta^*=\left[-\int_{0}^{1}\nabla^2L_{\con}(\theta^*+(\hat\theta-\theta^*)t)dt\right]^{-1}\nabla L_{\con}(\theta^*),
\end{equation*}
where the inverse exists by Assumption~\ref{ass:design}.
Next, we consider
\begin{align*}
	&\left\|J(\theta^*)+\int_{0}^{1}\nabla^2L_{\con}(\theta^*+(\hat\theta-\theta^*)t)dt\right\|_F\\
	\le&\Bigl\|\int_{0}^{1}\bigl[J(\theta^*)-J(\theta^*+(\hat\theta-\theta^*)t)\bigr]dt\Bigr\|_F\\
	&\quad+\Bigl\|\int_{0}^{1}\bigl[J(\theta^*+(\hat\theta-\theta^*)t)+\nabla^2L_{\con}(\theta^*+(\hat\theta-\theta^*)t)\bigr]dt\Bigr\|_F.
\end{align*}
Since $J(\theta)$ is continuous at $\theta^*$ and $\hat\theta\to_{\mathrm P}\theta^*$ by Theorem~\ref{thm:consistency}, the first term converges to zero in probability.
From Assumption~\ref{ass:clt1}, the second term also converges to zero in probability.
Therefore, we have
\begin{equation*}
	-\int_{0}^{1}\nabla^2L_{\con}(\theta^*+(\hat\theta-\theta^*)t)dt\to_{\mathrm P}J(\theta^*).
\end{equation*}
Note that $\sqrt M\,\nabla L_{\con}(\theta^*)=M^{-1/2}\sum_{i\in\cI}\xi_i$, where the summands are conditionally independent with mean zero, and its conditional covariance $-\nabla^2L_{\con}(\theta^*)\to_{\mathrm P}J(\theta^*)$ by Assumption~\ref{ass:clt1}.
Based on Assumption~\ref{ass:clt2}, the Lindeberg--Feller central limit theorem and the Cram\'er--Wold device give
\begin{equation*}
	\sqrt M\,\nabla L_{\con}(\theta^*)\to_{\mathrm d}\cN(0,J(\theta^*)).
\end{equation*}
From Slutsky's theorem, we have
\begin{equation*}
	\sqrt M(\hat\theta-\theta^*)=\left[-\int_{0}^{1}\nabla^2L_{\con}(\theta^*+(\hat\theta-\theta^*)t)dt\right]^{-1}\sqrt M\,\nabla L_{\con}(\theta^*)\to_{\mathrm d} \cN(0,J^{-1}(\theta^*)).
\end{equation*}
This completes the proof.
\qed

\subsection{Proof of Theorem~\ref{thm:support}}\label{asec:support}

We first recap the notation. The bandwidth is denoted as $h=h_n$.  We use $\epsilon:=\|\hat\theta-\theta^*\|_2$ to denote the convergence error in Theorem~\ref{thm:consistency}. The kernel function is given by $K_h(X):=h^{-d}K(X/h)$. For simplicity, we denote the debiased term as $\rho_i(\theta)=1-\bigl(1-\sigma(X_i^\top\theta)\bigr)^{m_i}$.
The oracle and plug-in reweighted outcomes are
\[
Z^*(X)=\sum_{i=1}^nw_i(X)Z_i^*,
\qquad
\hat Z(X)=\sum_{i=1}^nw_i(X)\hat Z_i,
\]
where
\[
w_i(X)=\frac{K_h(X_i-X)}{\sum_{j=1}^nK_h(X_j-X)},
\qquad
Z_i^*=\frac{\I(Y_i>0)}{\rho_i(\theta^*)},
\qquad
\hat Z_i=\frac{\I(Y_i>0)}{\rho_i(\hat\theta)}.
\]

\paragraph{Plug-in error.}
Note that for every $x\in(0,1)$
\begin{align*}
	1-(1-x)^{m}=x\sum_{k=0}^{m-1}(1-x)^{k}\ge m\,x(1-x)^{m-1},
\end{align*}
we have
\begin{equation*}\label{eq:grad_bound}
	0\le\frac{m\,x(1-x)^{m}}{1-(1-x)^{m}}
	\le1-x
	\le1.
\end{equation*}
By the mean value theorem, there exists $\tilde\theta$ between $\hat\theta$ and $\theta^*$ such that
\begin{equation}\label{eq:plug_bound}
	\bigl|\log\rho_i(\hat\theta)-\log\rho_i(\theta^*)\bigr|
	\le\left|\frac{m_i\,\sigma(X_i^\top\tilde\theta)\bigl(1-\sigma(X_i^\top\tilde\theta)\bigr)^{m_i}}{1-\bigl(1-\sigma(X_i^\top\tilde\theta)\bigr)^{m_i}}\right|\|X_i\|_2\|\hat\theta-\theta^*\|_2\le\epsilon\|X_i\|_2.
\end{equation}
According to Assumption~\ref{ass:kernel}, we have $w_i(X)=\I_{\cB(X,h)}(X_i)\cdot w_i(X)$. Since $\hat Z_i \rho_i(\hat\theta)= Z^*_i\rho_i(\theta^*)$, we have
\begin{equation*}
	\hat Z(X)=\sum_{i=1}^nw_i(X)\hat Z_i=\sum_{i=1}^nw_i(X) Z_i^*\I_{\cB(X,h)}(X_i) \exp(\log\rho_i(\theta^*)-\log\rho_i(\hat\theta))
\end{equation*}
Since $h<1$ when $n$ is sufficiently large, the Equation~\eqref{eq:plug_bound} implies
\[
e^{-\epsilon(\|X\|_2+1)}\I_{\cB(X,h)}(X_i)\le\I_{\cB(X,h)}(X_i) \exp(\log\rho_i(\theta^*)-\log\rho_i(\hat\theta))\le e^{\epsilon(\|X\|_2+1)}\I_{\cB(X,h)}(X_i).
\]
Furthermore, we have
\[
e^{-\epsilon(\|X\|_2+1)}Z^*(X)\le\hat Z(X)\le e^{\epsilon(\|X\|_2+1)}Z^*(X).
\]
\paragraph{Part (i).} We first show that kernel smoothing recovers a locally integrable function at its Lebesgue points. For a locally integrable $\psi$, we denote
\[
\psi_h(X):=\int_{\R^d}K_h(X'-X)\psi(X')\,dX'.
\]
By the Lebesgue differentiation theorem, almost every $X\in\R^d$ is a Lebesgue point of $\psi$. According to Assumption~\ref{ass:kernel}, every Lebesgue point $X$ of $\psi$ satisfies
\begin{equation}\label{eq:smoothing}
	|\psi_h(X)-\psi(X)|\le\frac{\bar\kappa}{h^d}\int_{\cB(X,h)}|\psi(X')-\psi(X)|\,dX'\to0.
\end{equation}

We then verify that $f(X)$, $f(X)\I_{S^*}(X)$ and $f(X)/\sigma(X^\top\theta^*)$ are locally integrable. By the model, $\P(Y_i>0\mid X_i,m_i)=\rho_i(\theta^*)\I_{S^*}(X_i)$ and $\rho_i(\theta^*)\ge\sigma(X_i^\top\theta^*)$, so
\[
\frac{1}{\sigma(X^\top\theta^*)}\le 1+e^{\|\theta^*\|_2\|X\|_2}.
\]
Since $f(X)$ is a density and $1/\sigma(X^\top\theta^*)$ is continuous, all three functions are locally integrable, and \eqref{eq:smoothing} applies to them. Fix $X$ with $f(X)>0$ that is a Lebesgue point of all three functions, which holds for almost every $X$ with $f(X)>0$. Let
\[
\hat f(X)=\frac1n\sum_{i=1}^nK_h(X_i-X),
\qquad
\hat g(X)=\frac1n\sum_{i=1}^nK_h(X_i-X)Z_i^*.
\]
Since $K_h^2\le h^{-d}\bar\kappa K_h$, Proposition~\ref{prop:Z} gives
\begin{align*}
	\E\hat f(X)&=\int_{\R^d}K_h(X'-X)f(X')\,dX', &
	\Var\hat f(X)&\le\frac{\bar\kappa}{nh^d}\int_{\R^d}K_h(X'-X)f(X')\,dX',\\
	\E\hat g(X)&=\int_{\R^d}K_h(X'-X)f(X')\I_{S^*}(X')\,dX', &
	\Var\hat g(X)&\le\frac{\bar\kappa}{nh^d}\int_{\R^d}K_h(X'-X)\frac{f(X')}{\sigma(X'^\top\theta^*)}\,dX'.
\end{align*}
By \eqref{eq:smoothing}, the means tend to $f(X)$ and $f(X)\I_{S^*}(X)$, and the variances are $O((nh^d)^{-1})$. Since $nh^d$ tends to infinity, Chebyshev's inequality gives
\[
\hat f(X)\to_{\mathrm P}f(X)>0,
\qquad
\hat g(X)\to_{\mathrm P}f(X)\I_{S^*}(X),
\qquad
Z^*(X)=\frac{\hat g(X)}{\hat f(X)}\to_{\mathrm P}\I_{S^*}(X).
\]
Since $\epsilon\to_{\mathrm P}0$, the plug-in error bound implies $\hat Z(X)\to_{\mathrm P}\I_{S^*}(X)$.
Let $\delta=\min\{\tau,1-\tau\}$. Since $\I_{S^*}(X)\in\{0,1\}$, we have 
\[
\I\{X\in\hat S\triangle S^*\}\le\I\{|\hat Z(X)-\I_{S^*}(X)|\ge\delta\}.
\]
By Tonelli's theorem,
\[
\E\,\mathcal{E}(\hat S)
=\int_{\R^d}\P\bigl(X\in\hat S\triangle S^*\bigr)f(X)\,dX
\le\int_{\R^d}\P\bigl(|\hat Z(X)-\I_{S^*}(X)|\ge\delta\bigr)f(X)\,dX.
\]
By dominated convergence and Markov's inequality, $\mathcal{E}(\hat S)\to_{\mathrm P}0$.

\paragraph{Part (ii).} Let $\delta=(1-\tau)/2$ and fix $\lambda>0$. We denote
\[
\mathcal A=\bigl\{\epsilon\le\lambda n^{-1/2}\bigr\},
\qquad
D=\Bigl\{X\in\R^d:\lambda n^{-1/2}(\|X\|_2+1)\le\log\frac{1+\tau}{2\tau}\Bigr\}.
\]
On $\mathcal A$, the plug-in error bound gives, for every $X\in D$,
\begin{equation}\label{eq:central}
	\hat Z(X)\ge\frac{2\tau}{1+\tau}Z^*(X).
\end{equation}

We first show that points outside $S^*\cup\partial_hS^*$ are never misclassified. Let $X\notin S^*\cup\partial_hS^*$. Since $\cB(X,h)$ is connected and does not meet $\partial S^*$, we have $\cB(X,h)\cap S^*=\emptyset$, and hence $Y_i\I_{\cB(X,h)}(X_i)=0$ almost surely. According to Assumption~\ref{ass:kernel},
\[
\hat Z(X)=\sum_{i=1}^nw_i(X)\hat Z_i\I_{\cB(X,h)}(X_i)=0,
\]
so $X\notin\hat S$.

We then bound the misclassification probability of points in $(S^*\cap D)\setminus\partial_hS^*$. For such $X$, $\cB(X,h)$ lies in the interior of $S^*$. We denote $N:=|\{i:X_i\in\cB(X,h/2)\}|$. According to Assumption~\ref{ass:kernel},
\[
w_i(X)\le\frac{\bar\kappa}{\kappa N}\qquad\text{when }N\ge1.
\]
By \eqref{eq:central}, on $\mathcal A$, $X\notin\hat S$ implies $Z^*(X)\le(1+\tau)/2=1-\delta$. Hence
\begin{equation}\label{eq:indicator}
	\I_{\mathcal A}\,\I\{X\in\hat S\triangle S^*\}\le\I\{N=0\}+\I\{N\ge1,\ Z^*(X)\le1-\delta\},
\end{equation}
whose right-hand side does not involve $\hat\theta$. Conditional on $\{(X_i,m_i)\}_{i=1}^n$, the $Y_i$ are independent and $\P(Y_i>0\mid X_i,m_i)=\rho_i(\theta^*)$ for $X_i\in\cB(X,h)$. Since $X_i^\top\theta^*\ge X^\top\theta^*-\|\theta^*\|_2$ and $\sigma(t-u)\ge e^{-u}\sigma(t)$ for $u\ge0$,
\[
\rho_i(\theta^*)\ge e^{-\|\theta^*\|_2}\sigma(X^\top\theta^*).
\]
Thus $w_i(X)Z_i^*$ are conditionally independent with total mean one and $0\le w_i(X)Z_i^*\le c$, where
\[
c=\frac{\bar\kappa\,e^{\|\theta^*\|_2}}{\kappa N\sigma(X^\top\theta^*)}.
\]
By the multiplicative Chernoff bound,
\[
\P\bigl(Z^*(X)\le1-\delta\bigm|\{(X_i,m_i)\}_{i=1}^n\bigr)\le e^{-\delta^2/(2c)}=e^{-aN},
\qquad
a=\frac{\kappa\delta^2e^{-\|\theta^*\|_2}\sigma(X^\top\theta^*)}{2\bar\kappa}\le1,
\]
and the bound also holds when $N=0$. Since $N\sim\mathrm{Bin}\bigl(n,\int_{\cB(X,h/2)}f(X')\,dX'\bigr)$ and $1-e^{-a}\ge a/2$, \eqref{eq:indicator} gives
\begin{align*}
	\P\bigl(\mathcal A\cap\{X\in\hat S\triangle S^*\}\bigr)
	&\le\E\,e^{-aN}
	=\Bigl\{1-(1-e^{-a})\int_{\cB(X,h/2)}f(X')\,dX'\Bigr\}^n\\
	&\le\exp\Bigl(-c_1n\,\sigma(X^\top\theta^*)\int_{\cB(X,h/2)}f(X')\,dX'\Bigr),
	\qquad
	c_1=\frac{\kappa\delta^2e^{-\|\theta^*\|_2}}{4\bar\kappa}.
\end{align*}

Finally, we integrate over $X$. By Tonelli's theorem and the two steps above,
\[
\E\bigl[\mathcal{E}(\hat S)\I_{\mathcal A}\bigr]
\begin{aligned}[t]
&\le\underbrace{\int_{\R^d\setminus D}f(X)\,dX}_{\mathrm{I}}
+\underbrace{\int_{\partial_hS^*}f(X)\,dX}_{\mathrm{II}}\\
&\quad+\underbrace{\int_{\R^d}\exp\Bigl(-c_1n\,\sigma(X^\top\theta^*)\int_{\cB(X,h/2)}f(X')\,dX'\Bigr)f(X)\,dX}_{\mathrm{III}}.
\end{aligned}
\]
By Markov's inequality, Assumption~\ref{ass:support_regular} and $nh\ge nh^d\to\infty$,
\[
\mathrm{I}\le\frac{\lambda^2\,\E(\|X\|_2+1)^2}{n\log^2\{(1+\tau)/(2\tau)\}}=O(h).
\]
By Assumption~\ref{ass:support_rate}(a), $\mathrm{II}\le Ch$. For $\mathrm{III}$, let
\[
s=\frac{2^d\log(1/h)}{c_1nh^d},
\]
which is at most one for large $n$. Outside $L_{h/2,s}$, the exponent is at most $-c_1ns(h/2)^d=-\log(1/h)$, so Assumption~\ref{ass:support_rate}(b) gives
\[
\mathrm{III}\le h+\int_{L_{h/2,s}}f(X)\,dX\le h+C(s^\alpha+h).
\]
Hence, with $C_\lambda$ depending on $\lambda$,
\[
\E\bigl[\mathcal{E}(\hat S)\I_{\mathcal A}\bigr]\le C_\lambda r_n,
\qquad
r_n=h+\Bigl(\frac{\log(1/h)}{nh^d}\Bigr)^{\alpha}.
\]
For any $b>0$, Markov's inequality gives
\[
\P\bigl(\mathcal{E}(\hat S)>br_n\bigr)\le\P(\mathcal A^c)+\frac{C_\lambda}{b}.
\]
Since $\epsilon=O_{\mathrm P}(n^{-1/2})$, choosing $\lambda$ and then $b$ large gives $\mathcal{E}(\hat S)=O_{\mathrm P}(r_n)$.
\qed

\end{document}